# INTERGLACIALS, MILANKOVITCH CYCLES, AND CARBON DIOXIDE

Gerald E. Marsh

Argonne National Laboratory (Ret)
5433 East View Park
Chicago, IL 60615

E-mail: gemarsh@uchicago.edu

**Abstract.** The existing understanding of interglacial periods is that they are initiated by Milankovitch cycles enhanced by rising atmospheric carbon dioxide concentrations. During interglacials, global temperature is also believed to be primarily controlled by carbon dioxide concentrations, modulated by internal processes such as the Pacific Decadal Oscillation and the North Atlantic Oscillation. Recent work challenges the fundamental basis of these conceptions.

**Introduction**

The history[1] of the role of carbon dioxide in climate begins with the work of Tyndall[2] in 1861 and later in 1896 by Arrhenius[3]. The conception that carbon dioxide controlled climate fell into disfavor for a variety of reasons until revived by Callendar[4] in 1938. It came into full favor after the work of Plass in the mid-1950s. Unlike what was believed then, it is known today that for Earth's present climate water vapor is the principal greenhouse gas with carbon dioxide playing a secondary role.

Climate models nevertheless use carbon dioxide as the principal variable while water vapor is treated as a feedback. This is consistent with, but not mandated by, the assumption that—except for internal processes—the temperature during interglacials is dependent on atmospheric carbon dioxide concentrations. It now appears that this is not the case: interglacials can have far higher global temperatures than at present with no increase in the concentration of this gas.

**Glacial Terminations and Carbon Dioxide**

Even a casual perusal of the data from the Vostok ice core shown in Fig. 1 gives an appreciation of how temperature[†] and carbon dioxide concentration change synchronously. The role of carbon dioxide concentration in the initiation of interglacials, during the transition to an interglacial, and its control of temperature during the interglacial is not yet entirely clear.

Between glacial and interglacial periods the concentration of atmospheric carbon dioxide varies between about 200-280 p.p.m.v, being at ~280 p.p.m.v. during interglacials. The details of the source of these variations is still somewhat controversial, but it is clear that carbon dioxide concentrations are coupled and in equilibrium with oceanic changes[5]. The cause of the glacial to interglacial increase in atmospheric carbon dioxide is now thought to be due to changes in ventilation of deep water at the ocean surface around Antarctica and the resulting effect on the global efficiency of the "biological pump".[6] If this is indeed true, then a perusal of the interglacial carbon dioxide concentrations show in Fig. 1 tell us that the process of increased ventilation coupled with an increasingly productive biological pump appears to be self-limiting during interglacials, rising little above ~280 p.p.m.v., despite warmer temperatures in past interglacials. This will be discussed more extensively below.

---

[†] Temperature is in terms of relative deviations of a ratio of oxygen isotopes from a standard. $^{18}O$ means:
$^{18}O = [((^{18}O/^{16}O)_{SAMPLE} - (^{18}O/^{16}O)_{STANDARD})/ (^{18}O/^{16}O)_{STANDARD})] \times 10^3$ ‰, measured in parts per thousand (‰). See: R. S. Bradley, *Paleoclimatology* (Harcourt Academic Press, New York 1999).

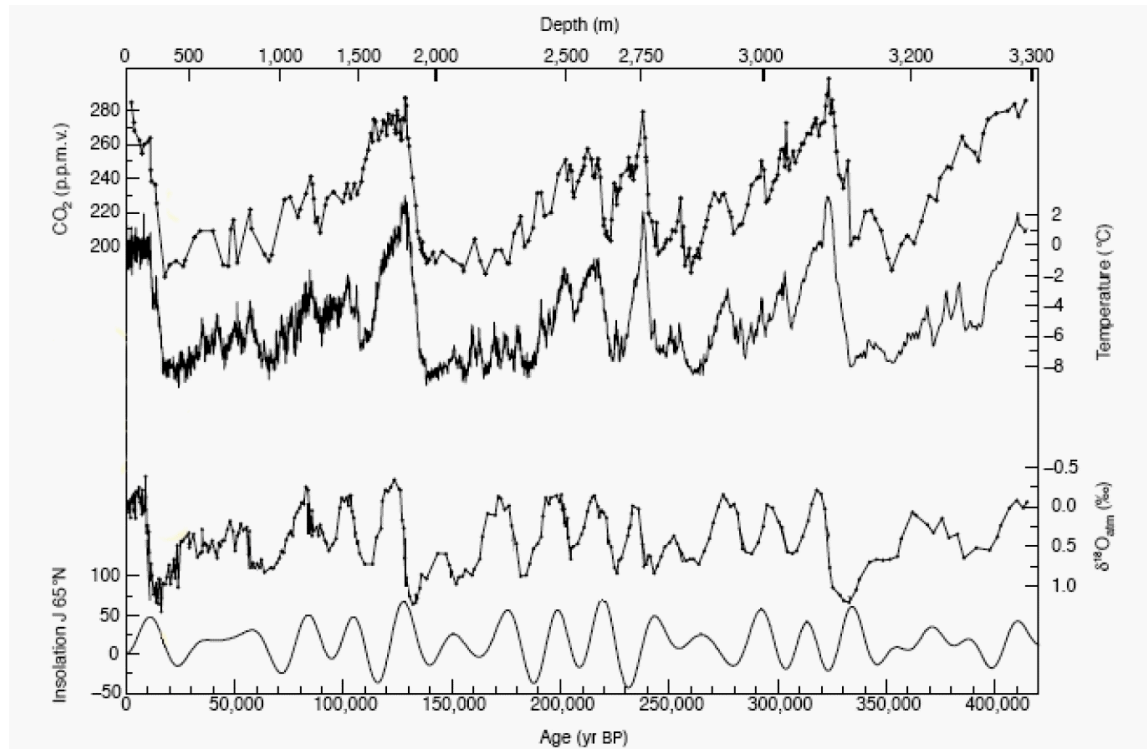

Fig. 1. Time series from the Vostok ice core showing CO2 concentration, temperature, $^{18}O_{atm}$, and mid-June insolation at 85°N in $Wm^{-2}$. Based on Fig. 3 of J. R. Petit, *et al.*[7]

The above mechanism for glacial to interglacial variation in carbon dioxide concentration is supported by the observation that the rise in carbon dioxide lags the temperature increase by some 800-1000 years—ruling out the possibility that rising carbon dioxide concentrations were responsible for terminating glacial periods. As a consequence, it is now generally believed that glacial periods are terminated by increased insolation in polar regions due to quasi-periodic variations in the Earth's orbital parameters. And it is true that paleoclimatic archives show spectral components that match the frequencies of Earth's orbital modulation. This Milankovitch insolation theory has a number of problems associated with it[8], and the one to be discussed here is the so called "causality problem"; i.e., what came first—increased insolation or the shift to an interglacial. This would seem to be the most serious objection, since if the warming of the Earth preceded the increased insolation it could not be caused by it.  This is not to say that Milankovitch variations in solar insolation do not play a role in changing climate, but they could not be the principal cause of glacial terminations.

Figure 2 shows the timing of the termination of the penultimate ice age (Termination II) some 140 thousand years ago.  The data shown is from Devils Hole (DH), Vostok, and the $^{18}O$ SPECMAP record.[†]  The DH-11 record shows that Termination II occurred at 140±3 ka; the Vostok record gives 140±15 ka; and the SPECMAP gives 128±3 ka.[9]  The latter is clearly not consistent with the first two.  The reason has to do with the origin of the SPECMAP time scale.

The SPECMAP record was constructed by averaging $^{18}O$ data from five deep-sea sediment cores.  The result was then correlated with the calculated insolation cycles over the last 800,000 years.  This procedure serves well for many purposes, but as can be seen from Fig. 2, sea levels were at or above modern levels before the rise in solar insolation often thought to initiate Termination II.  The SPECMAP chronology must therefore be adjusted when comparisons are made with records not dependent on the SPECMAP timescale.

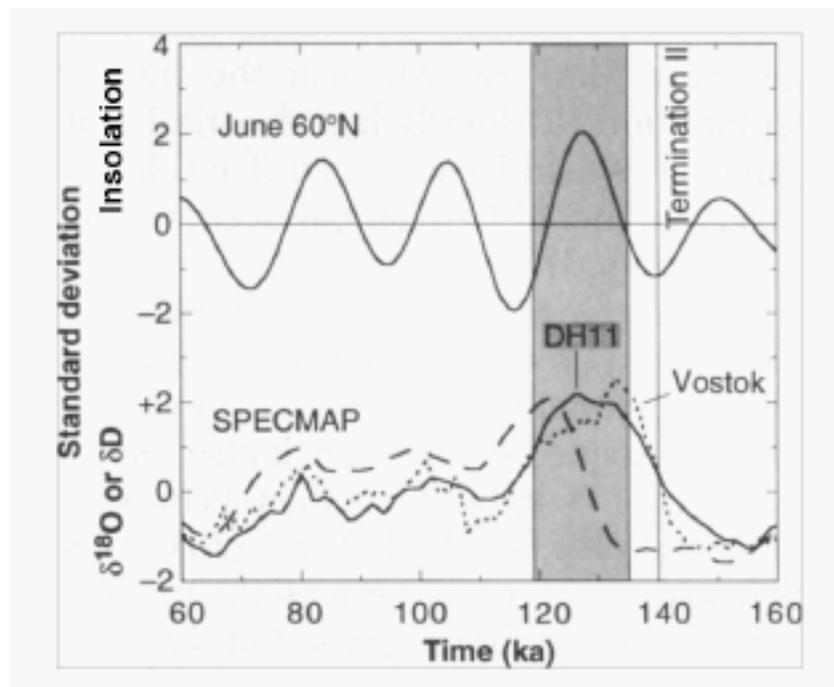

Fig. 2. Devils Hole (DH) is an open fault zone in south-central Nevada.  This figure shows a superposition of DH-11, Vostok, and SPECMAP curves for the period 160 to 60 ka in comparison with June 60°N insolation.  The shading corresponds to sea-levels at or above modern levels. [Fig 4 from: I. J. Winograd, *et al*., "Continuous 500,000-Year Climate Record from Vein Calcite in Devils Hole, Nevada", *Science* 258:255-260 (1992)]

---

[†] The acronym stands for Spectral Mapping Project.

The above considerations imply that Termination II was not initiated by an increase in carbon dioxide concentration or increased insolation.  The question then remains: What did initiate Termination II?

A higher time-resolved view of the timing of glacial Termination II is shown in Fig. 3 from Kirkby, *et al*.  What these authors found was that "the warming at the end of the penultimate ice age was underway at the minimum of 65°N June insolation, and essentially complete about 8 kyr prior to the insolation maximum".  In this figure, the Visser, *et al*. data and the galactic cosmic ray rate are shifted to an 8 kyr earlier time to correct for the SPECMAP time scale upon which they are based.  As discussed above, the SPECMAP timescale is tuned to the insolation cycles.

The galactic cosmic ray flux—using an inverted scale—is also shown in the figure.  The most striking feature is that the data strongly imply that Termination II was initiated by a reduction in cosmic ray flux.  Such a reduction would lead to a reduction in the amount of low-altitude cloud cover, as discussed below, thereby reducing the Earth's albedo with a consequent rise in global temperature[10].

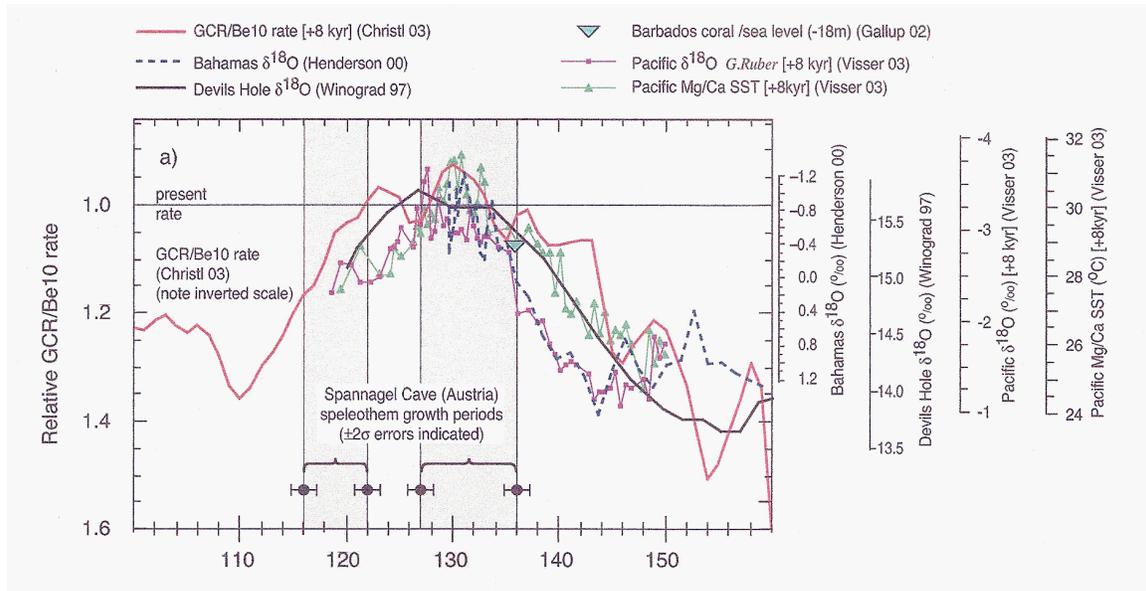

Figure 3. The timing of glacial Termination II. The figure shows the corrected galactic cosmic ray (GCR) rate[11] along with: the Bahamian $\delta^{18}$O record[12]; the date when the Barbados sea level was within 18 m of its present value[13] (Shown by the single inverted triangle at 136 kyr.); the $\delta^{18}$O temperature record from the Devils Hole cave in Nevada[14]; and the corrected Visser, *et al*. measurements of the Indo-Pacific Ocean surface temperature and $\delta^{18}$O records[15]. [From Kirkby, *et al*. (Ref. 19). The Winograd reference in the figure should be to 1992.]

There is another compelling argument that can be given to support this hypothesis. Sime, *et al*.[16] have found that past interglacial climates were much warmer than previously thought. Their analysis of the data shows that the maximum interglacial temperatures over the past 340 kyr were between 6 °C and 10 °C above present day values. From Fig. 1, it can be seen that past interglacial carbon dioxide concentrations were not higher than that of the current interglacial, and therefore carbon dioxide could not have been responsible for this warming. In fact, the concentration of carbon dioxide that would be needed to produce a 6-10 °C rise in temperature above present day values exceeds the maximum (1000 p.p.m.v.) for the range of validity of the usual formula [$\Delta F = \alpha \ln(C/C_0)$] used to calculate the forcing in response to such an increase.

In addition, it should be noted that the fact that carbon dioxide concentrations were not higher during periods of much warmer temperatures confirms the self-limiting nature of the process driving the rise of carbon dioxide concentration during the transition to interglacials; that is, where an increase in the ventilation of deep water at the surface of the Antarctic ocean and the resulting effect on the efficiency of the biological pump cause the glacial to interglacial rise carbon dioxide.

**Past Interglacials, Albedo Variations, and Cosmic Ray Modulation**

If it is assumed that solar irradiance during past interglacials was comparable to today's value (as is assumed in the Milankovitch theory), it would seem that the only factor left—after excluding increases in insolation or carbon dioxide concentrations—that could be responsible for the glacial to interglacial transition is a change in the Earth's albedo. During glacial periods, the snow and ice cover could not melt without an increase in the energy entering the climate system. This could occur if there was a decrease in albedo caused by a decrease in cloud cover.

The Earth's albedo is known to be correlated with galactic cosmic-ray flux. This relationship is clearly seen over the eleven-year cycle of the sun as shown in Fig. 4, which shows a very strong correlation between galactic cosmic rays, solar irradiance, and low cloud cover. Note that increased lower cloud cover (implying an increased albedo) closely follows cosmic ray intensity.

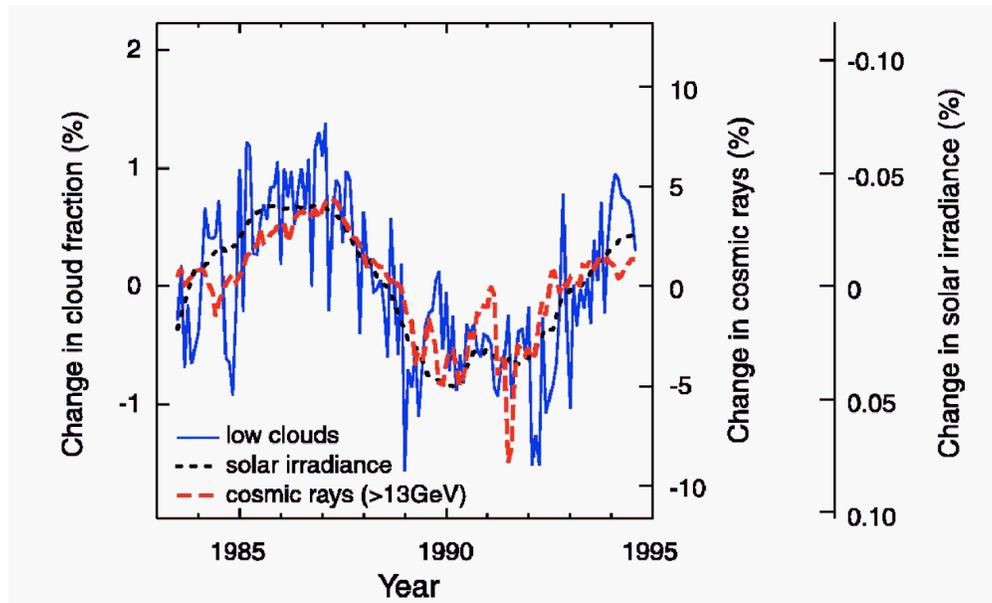

Figure 4. Variations of low-altitude cloud cover (less than about 3 km), cosmic rays, and total solar irradiance between 1984 and 1994. From K.S. Carslaw, R.G. Harrison, and J. Kirkby, *Science* **298**, 1732 (2002). Note the inverted scale for solar irradiance.

It is generally understood that the variation is galactic cosmic ray flux is due to changes in the solar wind associated with solar activity. The sun emits electromagnetic radiation and energetic particles known as the solar wind. A rise in solar activity—as measured by the sun spot cycle—affects the solar wind and the inter-planetary magnetic field by driving matter and magnetic flux trapped in the plasma of the local interplanetary medium outward, thereby creating what is called the heliosphere and partially shielding this volume, which includes the earth, from galactic cosmic rays—a term used to distinguish them from solar cosmic rays, which have much less energy.

When solar activity decreases, with a consequent small decrease in irradiance, the number of galactic cosmic rays entering the earth's atmosphere increases as does the amount of low cloud cover. This increase in cloud cover results in an increase in the earth's albedo, thereby lowering the average temperature. The sun's 11 year cycle is therefore not only associated with small changes in irradiance, but also with changes in the solar wind, which in turn affect cloud cover by modulating the cosmic ray flux. This, it is argued, constitutes a strong positive feedback needed to explain the significant impact of small changes in solar activity on climate. Long-term changes in cloud albedo would be associated with long-term changes in the intensity of galactic cosmic rays.

The great sensitivity of climate to small changes in solar activity is corroborated by the work of Bond, *et al*., who have shown a strong correlation between the cosmogenic nuclides $^{14}$C and $^{10}$Be and centennial to millennial changes in proxies for drift ice as measured in deep-sea sediment cores covering the Holocene time period[17]. The production of these nuclides is related to the modulation of galactic cosmic rays, as described above. The increase in the concentration of the drift ice proxies increases with colder climates[18]. These authors conclude that Earth's climate system is highly sensitive to changes in solar activity.

For cosmic ray driven variations in albedo to be a viable candidate for initiating glacial terminations, cosmic ray variations must show periodicities comparable to those of the glacial/interglacial cycles. The periodicities are shown in Fig. 5 taken (a) from Kirkby, *et* *al*.[19], and (b) from Schulz and Zeebe[20]. Figure 5(a) is derived from the galactic cosmic ray flux—shown in Fig. 6—over the last 220 kyr.

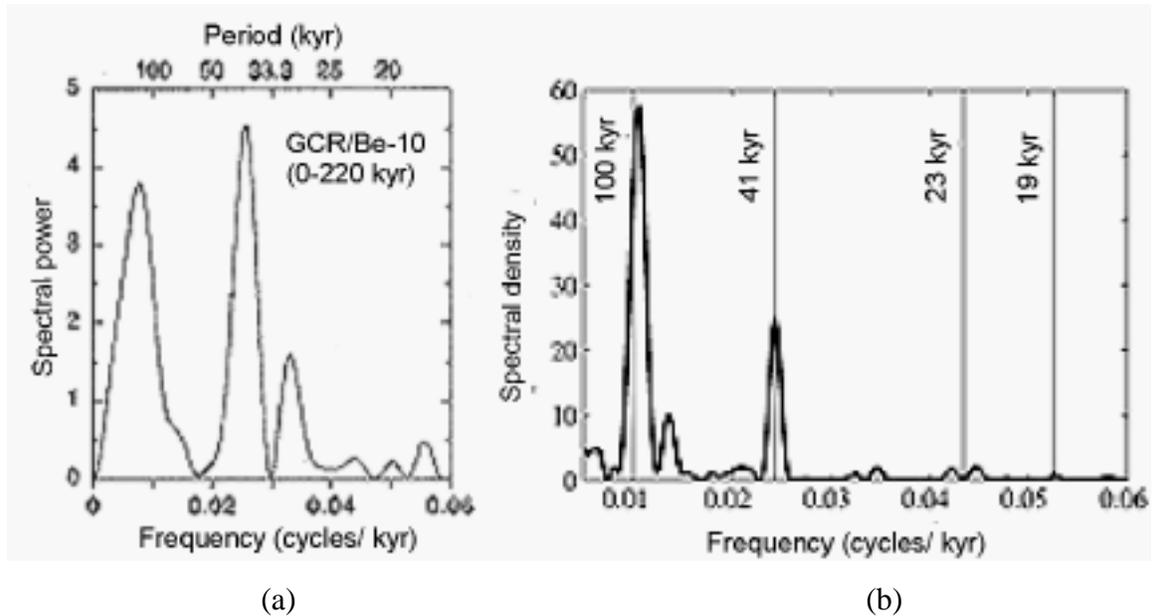

(a)            (b)

Figure 5. (a) Spectral power of the galactic cosmic ray flux for the past 220 ky as shown by $^{10}$Be in ocean sediments—from Fig. 6 below; (b) the de-trended and normalized $^{18}$O power spectral densities versus frequency over the last 900 kyr. [(a) from Kirkby, *et al*., and (b) from Schulz and Zeebe.] The resolution in (a) is not as good as in (b) because of the relatively short record.

The mechanism for the modulation of cosmic ray flux discussed above was tied to solar activity, but the 41 kyr and 100 kyr cycles seen in Fig. 5(a) correspond to the small quasiperiodic changes in the Earth's orbital parameters underlying the Milankovitch theory. For these same variations to affect cosmic ray flux they would have to modulate the geomagnetic field or the shielding due to the heliosphere. Although the existence of these periodicities and the underlying mechanism are still somewhat controversial, the lack of a clear understanding of the underlying theory does not negate the fact that these periodicities do occur in galactic cosmic ray flux.

If cosmic ray driven albedo change is responsible for Termination II, and a lower albedo was also responsible for the warmer climate of past interglacials—rather than higher carbon dioxide concentrations—the galactic cosmic ray flux would have had to be lower during past interglacials than it is during the present one. That this appears to be the case is suggested by the record in Fig. 6 (note inverted scale).

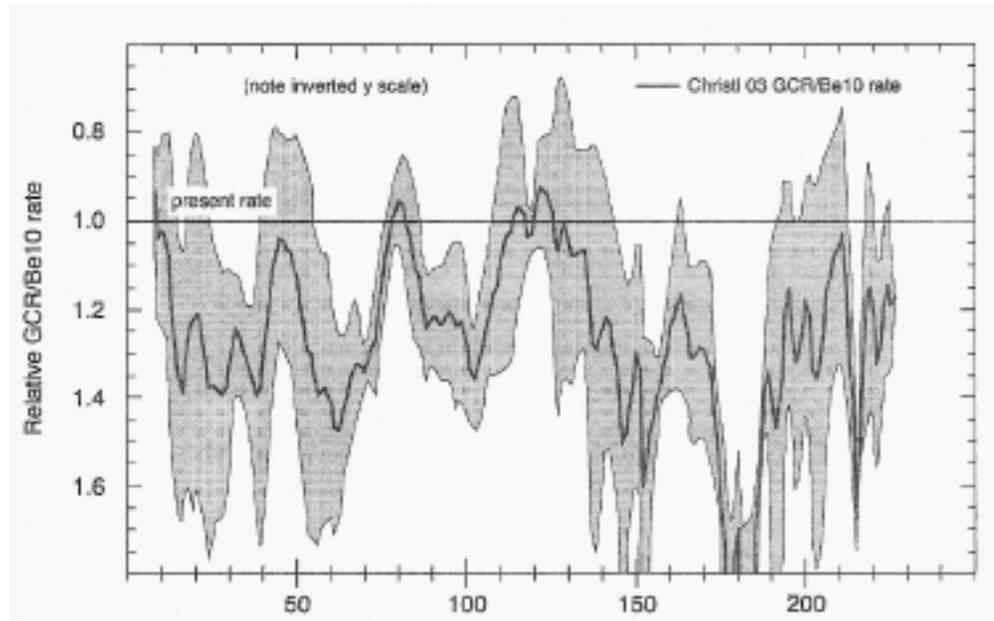

Figure 6. Galactic Cosmic Ray (GCR) flux over the last 220 kyr from combined $^{10}$Be deep-sea sediments. Based on the SPECMAP timescale. The shaded region represents the 1 confidence interval for the vertical scale. [From Kirkby, Mangine, and Muller (Ref. 19).]

Reconstructions of solar activity on the temporal scale of Fig. 6 are based on records of the cosmogenic radionuclides $^{14}$C and $^{10}$Be, and must be corrected for geomagnetic variations. Using physics-based models[21], solar activity can now be reconstructed over many millennia. The longer the time period, the greater the uncertainty due to systematic errors.

The appearance of the 100 ky and 41 ky periods in the galactic cosmic ray flux of Fig. 5(a) is somewhat surprising. With regard to orbital forcing of the geomagnetic field intensity, Frank[22] has stated that "Despite some indications from spectral analysis, there is no clear evidence for a significant orbital forcing of the paleointensity signals", although there were some caveats. In terms of the galactic cosmic ray flux, Kirby, et al.[23] maintain that ". . . previous conclusions that orbital frequencies are absent were premature". An extensive discussion of "Interstellar-Terrestrial Issues" has also been given by Scherer, et al.[24]

Using the fact that the galactic cosmic ray flux incident on the heliosphere boundary is known to have remained close to constant over the last 200 kyr, and that there exist independent records of geomagnetic variations over this period, Sharma[25] was able to use a functional relation reflecting the existing data to give a good estimate of solar activity over this 200 kyr period. The atmospheric production rate of $^{10}$Be depends on the geomagnetic field intensity and the solar modulation factor—the energy lost by cosmic ray particles traversing the heliosphere to reach the Earth's orbit (this is also known as the "heliocentric potential", an electric potential centered on the sun, which is introduced to simplify calculations by substituting electrostatic repulsion for the interaction of cosmic rays with the solar wind).

If $Q$ is the average production rate of $^{10}$Be, $M$ the geomagnetic field strength, and $\phi$ the solar modulation factor, the functional relation between the data used by Sharma is

$$Q(M_t, \phi_t) = Q(M_0, \phi_0) \left[ \frac{1}{a + b(\phi_t/\phi_0)} \right] \left[ \frac{c + (M_t/M_0)}{d + e(M_t/M_0)} \right],$$

(1)

where the subscript "0" corresponds to present day values. By setting these equal to unity, the normalized expression simplifies to

$$Q(M_t, \phi_t) = \left[ \frac{1}{a + b\phi_t} \right] \left[ \frac{c + M_t}{d + eM_t} \right].$$

(2)

The value of the constants[26] are: $a = 0.7476$, $b = 0.2458$, $c = 2.347$, $d = 1.077$, and $e = 2.274$. $Q$ may now be plotted as a function of both $M$ and $\phi$, as shown in Fig. 7.

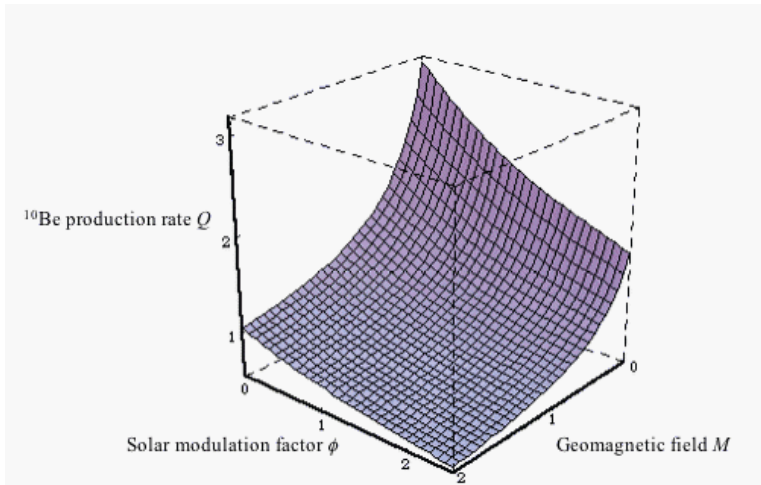

Figure 7. Atmospheric production rate of $^{10}$Be as a function of the solar modulation factor and the geomagnetic field intensity $M$. The plot corresponds to Eq. (2). The variables are normalized to present values of $Q$, $M$, and . Given the $^{10}$Be production rate and geomagnetic field from independent records, the normalized solar modulation factor may be calculated for any time in the past.

Using Eq. (2) and a globally stacked[†] record of relative geomagnetic field intensity from marine sediments, as well as a globally stacked, $^{230}$Th normalized $^{10}$Be record from deep marine sediments, Sharma was able to calculate the normalized solar modulation factor over the last 200 kyr. The result is shown in Fig. 8.

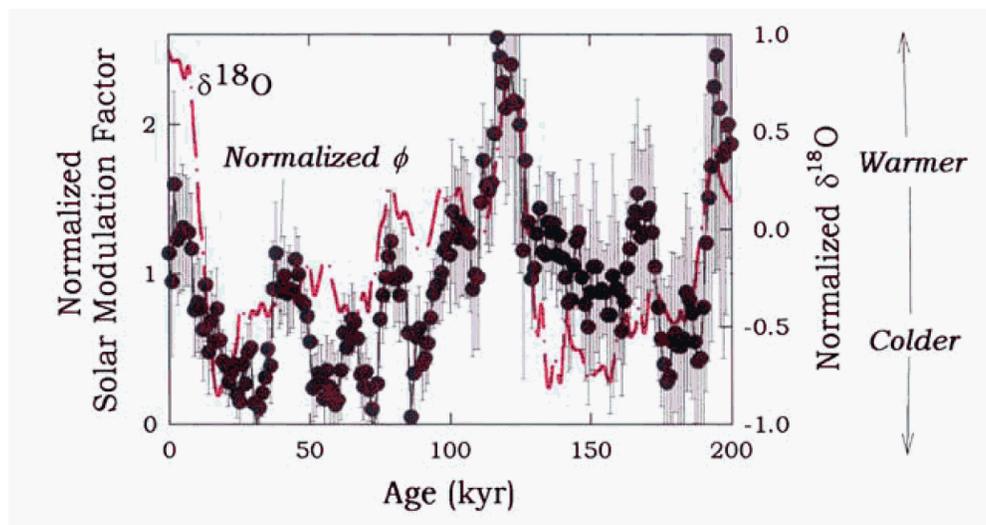

Figure 8. Normalized solar modulation factor and $^{18}$O record over the last 200 kyr [From Sharma (2002)]. The heavy dots correspond to values of the normalized solar modulation factor over the last 200 kyr. Each point is shown with an uncertainty of 1 standard deviation. Only the *y*-direction uncertainties are large enough to be seen in the plot.

---

[†] "Stacked" means aligned with respect to important stratigraphic features, but without a timescale.

The 100 kyr periodicity is readily apparent in Fig. 8. It is also seen that the $^{18}$O record and solar modulation are coherent and in phase. Sharma concludes from this that ". . . variations in solar surface magnetic activity cause changes in the Earth's climate on a 100-ka timescale".

Variations of solar activity over tens to hundreds of thousands of years do not seem to be a feature of the standard solar model. Ehrlich,[27] however, has developed a modification of the model based on resonant thermal diffusion waves that shows many of the details of the paleotemperature record over the last 5 million years. This includes the transition from glacial cycles having a 41 kyr period to a ~100 kyr period about 1 Myr ago. While the model has some problems—as noted by the author—the work shows that a reasonable addition to the standard solar model could help explain the long-term paleotemperature record. However, the fact that ice ages did not begin until some 2.75 million years ago, when atmospheric carbon dioxide concentration levels fell to values comparable to today, tells us that solar variations consistent with Ehrlich's model did not significantly affect climate before this oscillatory phase of climate began[28].

**Summary**

It has been shown above that low altitude cloud cover closely follows cosmic ray flux; that the galactic cosmic ray flux has the periodicities of the glacial/interglacial cycles; that a decrease in galactic cosmic ray flux was coincident with Termination II; and that the most likely initiator for Termination II was a consequent decrease in Earth's albedo.

The temperature of past interglacials was higher than today most likely as a consequence of a lower global albedo due to a decrease in galactic cosmic ray flux reaching the Earth's atmosphere. In addition, the galactic cosmic ray intensity exhibits a 100 kyr periodicity over the last 200 kyr that is in phase with the glacial terminations of this period. Carbon dioxide appears to play a very limited role in setting interglacial temperature.

**ENDNOTES AND REFERENCES**